\documentclass[conference]{IEEEtran}

\usepackage{fancyhdr}
\IEEEoverridecommandlockouts
\usepackage{algorithm}
\usepackage{algpseudocode}
\usepackage{amsfonts}
\usepackage{bm}
\usepackage{amsmath}
\usepackage{cite}
\usepackage{float}
\usepackage{amssymb}
\usepackage{enumerate}
\usepackage{color,xcolor}
\usepackage{graphicx}
\usepackage{multirow,tabularx}
\usepackage{subfigure}
\usepackage{hyperref}

\makeatletter

\newcommand{\Rmnum}[1]{\expandafter\@slowromancap\romannumeral #1@}
\makeatother
\usepackage{bookmark}

\ifCLASSINFOpdf

\else

\fi

\begin{document}

\title{{Gaussian Message Passing Iterative Detection for MIMO-NOMA Systems with Massive Access}}

\author{\IEEEauthorblockN{Lei Liu\IEEEauthorrefmark{1}, Chau Yuen\IEEEauthorrefmark{2}, Yong Liang Guan \IEEEauthorrefmark{3}, Ying Li\IEEEauthorrefmark{1} and Chongwen Huang\IEEEauthorrefmark{2}\\
\IEEEauthorrefmark{1}State Key Lab of ISN, Xidian University, Xi'an 710071, China\\
\IEEEauthorrefmark{2}Singapore University of Technology and Design, Singapore\\
\IEEEauthorrefmark{3}Nanyang Technological University, Singapore\\
E-mail:yli@mail.xidian.edu.cn}

%\thanks{This work was supported in part by the 973 Program under Grant 2012CB316100, the National Natural Science Foundation of China under Grants 61301177 and 61550110244, and Singapore A*STAR SERC Project under Grant 142 02 00043. The first author was also supported by the China Scholarship Council under Grant 20140690045. }
}

%\author{\IEEEauthorblockN{Lei Liu\IEEEauthorrefmark{1}, Ying Li\IEEEauthorrefmark{2}, Chau Yuen\IEEEauthorrefmark{3} and Yongliang Guan \IEEEauthorrefmark{4}\\
%\IEEEauthorrefmark{1}\IEEEauthorrefmark{4}State Key Lab of ISN, Xidian University, Xi'an, China\\
%\IEEEauthorrefmark{2}Singapore University of Technology and Design, Singapore\\
%\IEEEauthorrefmark{3}Nanyang Technological University, Singapore\\
%Email: \IEEEauthorrefmark{1}lliu\_0@stu.xidian.edu.cn,
%\IEEEauthorrefmark{2}yli@mail.xidian.edu.cn,\\
%\quad\;\IEEEauthorrefmark{3} yuenchau@sutd.edu.sg,
%\IEEEauthorrefmark{4}eylguan@ntu.edu.sg
%}}

\maketitle

\begin{abstract}
This paper considers a low-complexity Gaussian Message Passing Iterative Detection (GMPID) algorithm for Multiple-Input Multiple-Output systems with Non-Orthogonal Multiple Access (MIMO-NOMA), in which a base station with $N_r$ antennas serves $N_u$ sources simultaneously. Both $N_u$ and $N_r$ are very large numbers and we consider the cases that $N_u>N_r$. The GMPID is based on a fully connected loopy graph, which is well understood to be not convergent in some cases. The large-scale property of the MIMO-NOMA is used to simplify the convergence analysis. Firstly, we prove that the variances of the GMPID definitely converge to that of Minimum Mean Square Error (MMSE) detection. Secondly, two sufficient conditions that the means of the GMPID converge to a higher MSE than that of the MMSE detection are proposed. However, the means of the GMPID may still not converge when $ N_u/N_r< (\sqrt{2}-1)^{-2}$. Therefore, a new convergent SA-GMPID is proposed, which converges to the MMSE detection for any $N_u> N_r$ with a faster convergence speed. Finally, numerical results are provided to verify the validity of the proposed theoretical results.
\end{abstract}

%\begin{IEEEkeywords}
%MIMO-NOMA, Gaussian message passing, Massive access, Convergence, low-complexity detection.
%\end{IEEEkeywords}

\IEEEpeerreviewmaketitle
\section{Introduction}
 The demand of wireless communication has increased exponentially with the fast development of communication theory in these years. It is forecasted that the number of wireless communication devices will reach 40.9 billion and the units of Internet of Things (IoT) will increase to 26 billion by 2020 \cite{Gartner, ABI, Tushar2016}. Therefore, due to the limited spectrum resources, the massive access in the same time/frequency is inevitable in the future wireless communication systems \cite{Lien, FuHui2016}. All the users are allowed to be superimposed in the same time/code/frequency domain in \emph{Non-Orthogonal Multiple Access} (NOMA) to significantly increase the spectral efficiency and reduce latency in the future communication systems \cite{ Ding2014,  Kim2015, Lei2016}. In addition, the system performance can be further enhanced by combining NOMA with the \emph{Multiuser Multiple-Input and Multiple-Output} (MU-MIMO)\cite{ Lei2016}, which is a key technology for wireless communication standards like IEEE 802.11 (Wi-Fi), WiMAX and Long Term Evolution (4G) \cite{ Rusek2013, JunZuo2015, HuiGao2015, Kong2016, Wei2015, Lei2015, Lei20162}. In the massive access MIMO systems, the number of users $N_u$ is larger than the number of antennas $N_r$, i.e., $N_u\!\!>\!\!N_r$, which is much different with the massive MIMO that requires $N_u\!\!<\!\!<\!\!N_r$ \cite{Rusek2013}.

Unlike the \emph{Orthogonal Multiple Access} systems, the signal processing in the NOMA systems will cost a higher complexity at the \emph{base station} (BS)\cite{Rusek2013}. Low-complexity uplink detection for MIMO-NOMA is a challenging problem due to the interference between the users \cite{Rusek2013, Kim2015}, especially for the massive access. In \cite{Lei2016}, it shows that with proper designed superposition coded modulation (SCM), Iterative Linear Minimum Mean Square Error (LMMSE) detection approaches the sum capacity of the MIMO-NOMA systems. However, the complexity of LMMSE detection is high due to large matrix inversion \cite{tse2005}. To avoid the matrix inversion, the classical iterative algorithms like Jacobi and Richardson algorithm are applied in \cite{Axelsson1994, Gao2014}. Another promising detection for MIMO-NOMA is message passing algorithm (MPA) \cite{ Forney2001,  Loeliger2004}. There are two MPAs, one of which is the Gaussian Belief Propagation (GaBP) algorithm based on a graph consisted of variable nodes \cite{  malioutov2006, Su2015}, the other one is the Gaussian Message Passing Iterative Detection (GMPID) based on a pairwise graph consisted of variable nodes and sum nodes \cite{ andrea2005, Roy2001, wu2014,Lei2015}.

For the factor graph with a tree structure, the means and variances of the message passing algorithm converge to the true marginal means and variances respectively \cite{Loeliger2004}. However, if the graph has cycles, the message passing algorithm may fail to converge. There are three well known sufficient convergence conditions of the loopy GaBP algorithm, i.e., diagonal-dominance, convex decomposition and walk-summability \cite{malioutov2006}. Recently, a necessary and sufficient convergence condition of the GaBP is given in \cite{Su2015}. For the GMPID based on the pairwise graph, a sufficient condition of the mean convergence is given in \cite{Roy2001}. However, the high-complexity matrix inversion is calculated at the sum nodes during the message updating. In general, the GMPID has lower computational complexity and a better Mean Square Error (MSE) performance than the GaBP algorithm \cite{Lei2015}. Montanari \cite{andrea2005} has proved that the GMPID converges to the optimal MMSE solution for any arbitrarily loaded randomly-spread CDMA system, but it only works for CDMA MU-MIMO system with binary channels. In \cite{Lei2015}, the convergence of GMPID based on a pairwise graph is analysed and improved for the massive MIMO that $N_u<N_r$. However, the convergence of GMPID for the MIMO-NOMA that $N_u>N_r$ still remains unknown.

In this paper, based on the large scale property of the system (large $N_u$ and $N_r$), we analyse the convergence of the GMPID and propose a new fast-convergence detector for massive access MIMO-NOMA. Let $\beta  = {N_u \mathord{\left/
 {\vphantom {N_u N_r}} \right.
 \kern-\nulldelimiterspace} N_r}$ and $\beta>1$. The contributions of this paper are summarized as follows.

 \noindent
\hangafter=1
\setlength{\hangindent}{2em} 1) We prove that the variances of GMPID definitely converge to the MSE of MMSE detection.

 \noindent
\hangafter=1
\setlength{\hangindent}{2em} 2) Two sufficient conditions that the means of GMPID converge to a higher MSE performance than that of the MMSE detector for $\{\beta:\! \beta\!>\!(\!\sqrt{2}\!-\!1)^{-2}\}$ are derived.

 \noindent
\hangafter=1
\setlength{\hangindent}{2em} 3) A new fast-convergence detector called SA-GMPID is proposed, which converges to the MMSE detection with a faster convergence speed for any $\{\beta: \beta>1\}$.\vspace{-0.1cm}

\section{System Model}
In this section, the MIMO-NOMA system model and some preliminaries about the iterative detection for the massive MIMO-NOMA systems are introduced.
\subsubsection{System Model}
Consider an uplink MU-MIMO system as showed in Fig \ref{f2}. In this system, $N_{u}$ autonomous single-antenna terminals simultaneously communicate with an array of $N_{r}$ antennas of the base station (BS) under the same frequency. We consider the system has massive users, i.e., $N_{u}$ is large and $N_{u}>N_{r}$. All the users interfere with each other at the receiver and are non-orthogonal both in time domain and frequency domain in the NOMA systems due to the large number of users. Then, the $N_{r}\times1$ received ${ \mathbf{y}_t}$ at time $t$ is\vspace{-0.15cm}
\begin{equation}\label{e1}
{ \mathbf{y}_t}= \mathbf{H}{ \mathbf{x}^{tr}_t} + \mathbf{n}_t,\quad t\in \mathcal{N},\vspace{-0.15cm}
\end{equation}
where $\mathcal{N}=\{1,\cdots,N\}$, $\mathbf{H}$ is a $N_{r} \times N_{u}$ channel matrix, $\mathbf{n}_t\sim\mathcal{CN}^{{N_{r}}}(0,\sigma_n^2)$ is the $N_{r} \times 1$ independent additive white Gaussian noise (AWGN) vector at time $t$, and $\mathbf{x}^{tr}_t=[x_{1,t},\cdots,x_{N_u,t}]^T$ is the message vector sent from $N_{u}$ users. In this paper, we consider the fading channels $\mathbf{H}$ whose entries are i.i.d. with $\mathcal{CN}(0,1)$, and the BS knows $\mathbf{H}$.
\subsubsection{Transmitters}
\begin{figure}[t]\vspace{-0.4cm}
  \centering
  \includegraphics[width=9.2cm]{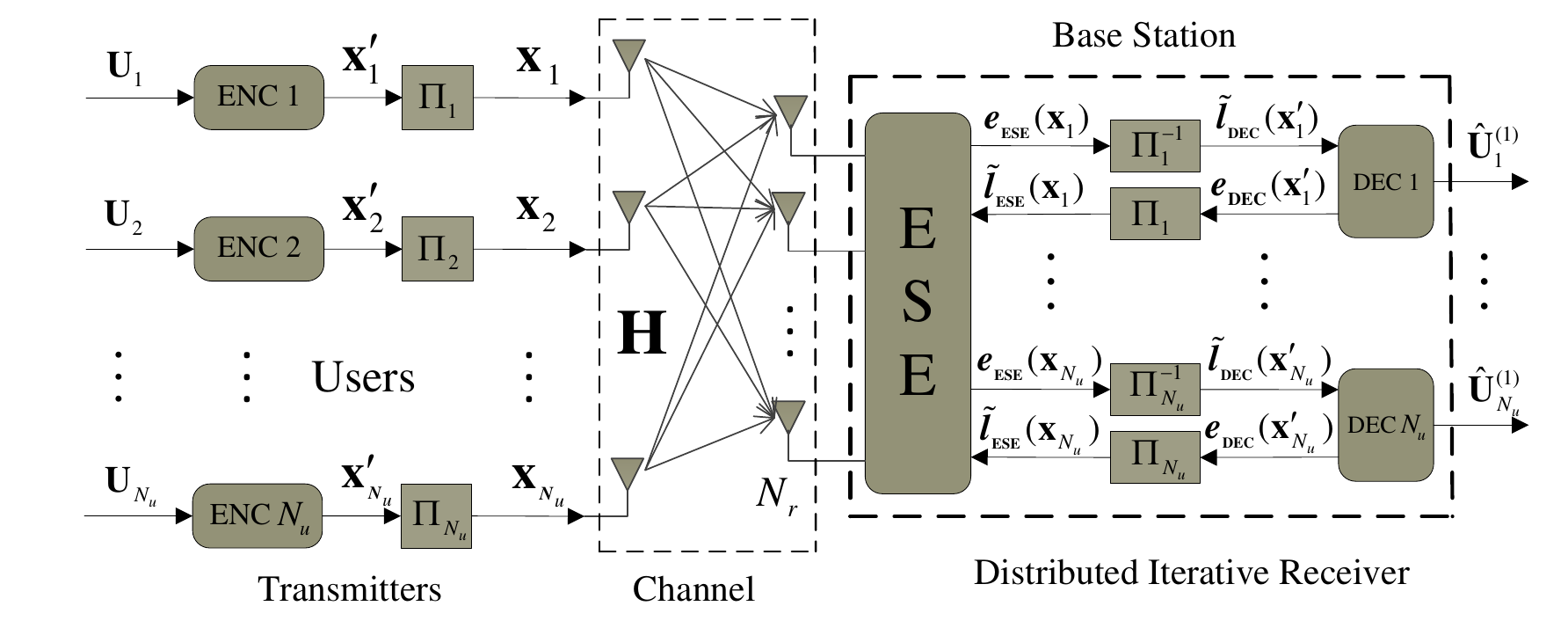}\\\vspace{-0.35cm}
  \caption{Block diagram of the MIMO-NOMA system massive access: the number of users $N_u$ is larger than the number of antennas $N_u$. ENC is the encoder and DEC is the decoder. $\Pi_i$ denotes the interleaver and $\Pi_i^{-1}$ denotes the de-interleaver. ESE represents the elementary signal estimator. $\mathbf{H}$  contains the small-scale fading channels. }\label{f2}\vspace{-0.5cm}
\end{figure}

As illustrated in Fig. \ref{f2}, at user $i$, an information sequence ${\bf{U}}_i $ is encoded by a channel code with rate $R_i$ into a $N$-length coded sequence ${\mathbf{x}}'_{i}$, $\mathop{i}\in \mathcal{N}_{u},\; \mathcal{N}_{u}= \left\{ {{{1,2,}} \cdots {{,N_{u}}}} \right\}$ and then interleaved by an {$N$-length} independent random interleaver $\Pi_{i}$ and get $\mathbf{x}_{i}=[x_{i,1},x_{i,2},\cdots,x_{i,N}]^T$. In this paper, we assume that $\mathbf{x}_{i}\sim \mathcal{N}^N(0,\sigma^2_{x_i})$ for $i\in \mathcal{N}_u$.  {\footnote{This assumption does not lose any generality since signal shaping to approach Gaussian signaling can be realized by properly designing the constellation points of the transmissions. According to the Shannon theory \cite{Cover2006}, it is well known that the capacity of Gaussian channel is achieved by a Gaussian input. Therefore, the independent Gaussian sources assumption are widely used in the communication networks \cite{Kafedziski2012}.}}

\subsubsection{Iterative Receiver}
At the base station, the received signals $\mathbf{Y}=[\mathbf{y}_1,\cdots,\mathbf{y}_N]$ and message  $\{{\tilde {\textbf{\emph{l}}}_{\small{ESE}}} ({\mathbf{x}}_{i}),i\in \mathcal{N}_u\}$ from the decoder are sent to a low-complexity elementary signal estimator (ESE) to estimate the extrinsic message ${\large{\textbf{\emph{e}}}_{\small{ESE}}}({\mathbf{x}}_{i})$ for decoder $i$, which is then deinterleaved with $\Pi_i^{-1}$ into ${\tilde {\textbf{\emph{l}}}_{\small{DEC}}} ({\mathbf{x}}'_{i})$, $i\in \mathcal{N}_u$. The corresponding single-user decoder employs ${\tilde {\textbf{\emph{l}}}_{\small{DEC}}} ({\mathbf{x}}'_{i})$ as the prior message to calculate the extrinsic message ${\large{\textbf{\emph{e}}}_{\small{DEC}}}({\mathbf{x}}'_{i})$. Similarly, this extrinsic message is interleaved by $\Pi_i$ to obtain the prior information ${\tilde {\textbf{\emph{l}}}_{\small{ESE}}} ({\mathbf{x}}_{i})$ for the ESE. Repeat this process until the maximum number of iteration is achieved or the MSE meets the requirement.  Actually, the messages ${\large{\textbf{\emph{e}}}_{\small{ESE}}}({\mathbf{x}}_{i})$, and ${\tilde {\textbf{\emph{l}}}_{\small{ESE}}} ({\mathbf{x}}_{i})$  can be replaced by $[\bar{\mathbf{x}}^e_i, \mathbf{v}^e_{\bar{\mathbf{x}}_i}]$ and $[\bar{{\mathbf{x}}}^l_i, \mathbf{v}^l_{\bar{\mathbf{x}}_i}]$ respectively for any $i\in\mathcal{N}_u$, if the messages are all Gaussian distributed.
\subsubsection{Element Signal Estimator (ESE)}
For the ESE, with the input prior messages ${\tilde {\emph{{{l}}}}_{\small{ESE}}} ({{x}}_{i,t})$, we can calculate the prior mean and variance of $x_i(t)$ by\vspace{-0.15cm}
\begin{equation}
{{\bar x}_{i,t}^l} = E\left[{x_{i,t}}|{\tilde {\emph{{{l}}}}_{\small{ESE}}} ({{x}}_{i,t})\right],\quad\bar{v}^l_i = E\left[|{x_{i,t}}-\bar{x}^l_{i,t}|^2\right],\vspace{-0.15cm}
\end{equation}
for any index $t$. Actually, $\bar{v}_i^l$ is time invariant. Then, the input messages ${\tilde {\emph{{{l}}}}_{\small{ESE}}} ({{x}}_{i,t})$ are replaced by the ${{\bar {{x}}}_{i,t}^l}$ and $\bar{v}_i^l$.

The extrinsic mean and variance for $x_{i,t}$ (denoted by $u_i$ and $b_{i,t}$) are calculated according to the Gaussian message combining rule \cite{Loeliger2004} by\vspace{-0.15cm}
\begin{equation}\label{e9}
\bar{v}_i^{e^{-1}} = \hat{v}_{i}^{-1}-\bar{v}_i^{l^{-1}}\;
\mathrm{and} \;
\frac{\bar{x}_{i,t}^e}{\bar{v}_i^{e} }=\frac{\hat{x}_{i,t}}{\hat {v}_{i}}-\frac{\bar{x}^l_{i,t}}{\bar{v}_i^{l}},\vspace{-0.15cm}
\end{equation}
where $i\in \mathcal{N}_u$, $t\in \mathcal{N}$, and $\hat{v}_i$ and $\hat{x}_{i,t}$ are the variance and mean of the estimation (total messages that contain the prior and extrinsic messages) for $x_{i,t}$.

\section{LMMSE ESE and Performance Estimation}
For the gaussian sources, LMMSE detection is an optimal linear detector under MSE measure as it minimizes the MSE between sources and estimation \cite{verdu1998}. In addition, it is proved that iterative LMMSE detector achieves the sum capacity of MIMO-NOMA with proper channel code design \cite{Lei2016}.

Let $\bar{\mathbf{x}}^l_t=[{\bar{x}_{1,t}^l},\cdots,\bar{x}_{N_u,t}^l]$, $\mathbf{V}^l_{\bar{\mathbf{x}}_t}=\mathbf{V}^l_{\bar{\mathbf{x}}}= \mathrm{diag}(\bar{v}_1^l,\bar{v}_2^l,\cdots,\bar{v}_{N_u}^l)$, and $\mathbf{V}_{\mathbf{x}}$ denote the covariance matrix of the sources $\mathbf{x}$. The output of the LMMSE detector \cite{tse2005} is \vspace{-0.15cm}
\begin{equation}\label{GMP2}
{{\hat {\mathbf{x}}}_t} = \mathbf{V}_{\hat {\mathbf{x}}}\left[\mathbf{V}_{\bar{\mathbf{x}}}^{l^{-1}}\bar{\mathbf{x}}_t^l+ \sigma^{-2}_n\mathbf{H}^H\mathbf{y}_t  \right],\vspace{-0.15cm}
\end{equation}
where $\mathbf{V} _{{{\hat {\mathbf{x}}}}} = (\sigma _{{{n}}}^{- 2}\mathbf{H}^H\mathbf{H}+\mathbf{V} _{{{ \bar{\mathbf{x}}}}}^{l^{-1}})^{-1}$.

\textbf{\textit{Proposition 1:}} \cite{verdu1998} \emph{When $N_u,N_r\rightarrow \infty$ with $\beta= N_u/N_r$ and $\mathbf{V}_{\bar{\mathbf{x}}}^l=\bar{v}^lI_{N_u}$, the MSE performance of the LMMSE detection for the symmetric MIMO-NOMA systems is\vspace{-0.05cm}
\begin{equation*}\label{PA6}
\!\!\hat{v}_{mmse}\!\!= \!\bar{v}^l \!- \frac{\sigma_n^2}{{4N_u}}\Big(\!\!\sqrt {\!\!snr^lN_r{{( {1\! + \!\sqrt \beta } )}^2} \!\!\!+\! 1} -\sqrt {\!\!snr^lN_r{{( {1 \!-\! \!\sqrt \beta })}^2} \!\!\!+\!\! 1}  \Big)^2\!\!,\vspace{-0.15cm}
\end{equation*}
where $i\in {\mathcal{N}}_u$ and $snr^l={{\bar{v}^l} \mathord{\left/
 {\vphantom {{\bar{v}^l} {\sigma _n^2}}} \right.
 \kern-\nulldelimiterspace} {\sigma_n^2}}$ is the signal-to-noise ratio.}

For the uncoded symmetric MIMO-NOMA systems with fixed $\bar{v}^l=\sigma_x^2$, the performance of the uncoded MIMO-NOMA system is very poor when $\beta>1$, because the interference between the users limits the system performance \cite{Lei2015}. Hence, the channel code for each user is used to combat the user interference in the MIMO-NOMA systems. In the coded MIMO-NOMA system, the variance $\bar{v}^l$ decreases with the number of iterations. As a result, the system performance is improved through the joint iteration between the LMMSE detector and user decoders.

The complexity of LMMSE detection is $\mathcal{O}((N_u^3+N_rN_u^2)N_{ite}^{out})$, where $N_{ite}^{out}$ is the number of out iterations between the ESE and decoders. However, the complexity of LMMSE estimator is still too high when the number of users is very large.

\section{Gaussian Message Passing Iterative Detector}
\begin{figure}[t]\vspace{-0.4cm}
  \centering
  \includegraphics[width=9.0cm]{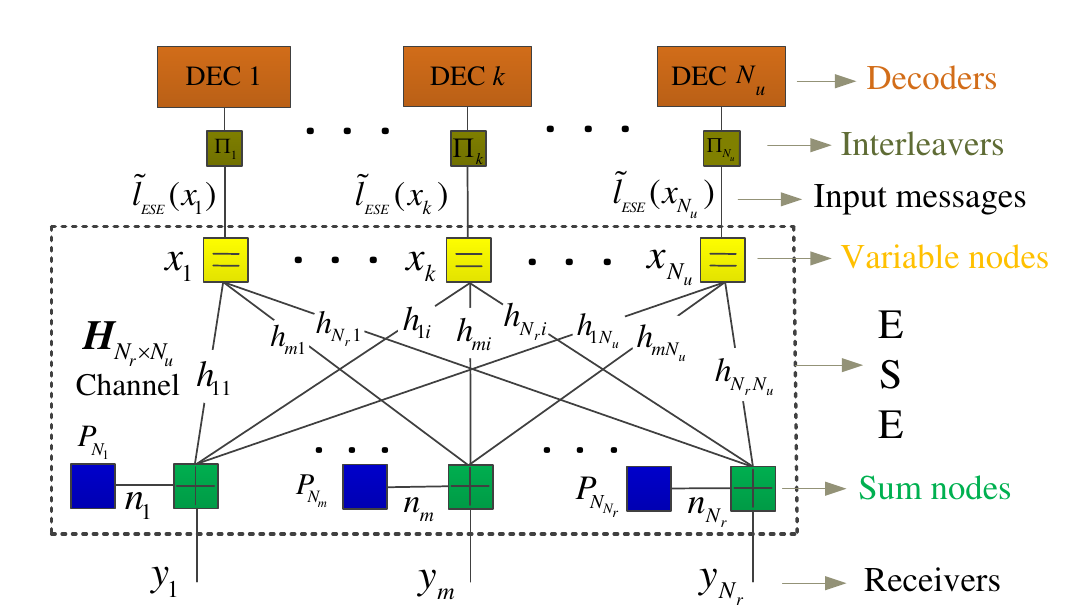}\\\vspace{-0.35cm}
  \caption{GMIPD for coded MIMO-NOMA systems.} \label{fac_graph} \vspace{-0.45cm}
\end{figure}

In Fig. \ref{fac_graph}, the GMPID based on a pairwise factor graph is considered for the MIMO-NOMA systems. However, the GMPID that based on a fully connected loopy factor graph does not always converge. In \cite{Lei2015}, for the MU-MIMO system with $N_u<N_r$, the convergence analysis of the GMPID is given. However, for the case $N_u>N_r$, the results in \cite{Lei2015} are not suitable any more, and the convergence analysis is more difficult due to intractable interference between the large number of users. In this section, for the coded MIMO-NOMA with $N_u>N_r$, the GMPID and its convergence analysis will be proposed. Due to the chip-by-chip ESE, we drop the subscript $t$ in the rest of this paper.

\subsection{Message Update at Sum Nodes of GMPID}
The message update at the sum nodes is given by \vspace{-0.2cm}
\begin{equation}\label{np1}
\left\{ \begin{array}{l}
x_{m \to k}^s(\tau) = {y_m} - \sum\limits_{i} {h_{mi}x_{i \to m}^v(\tau-1)\;,} \\
v_{m \to k}^s(\tau)= \sum\limits_{i} {h_{mi}^2v_{i \to m}^v(\tau-1) + \sigma _n^2\;},
\end{array} \right.\vspace{-0.2cm}
\end{equation}
where $i,k \in \mathcal{N}_u ,m \in \mathcal{N}_r$, $\tau$ denotes the $\tau$th iteration, $y_m$ is the $m$th element of the received vector $\textbf{\emph{y}}$, and $h_{mi}$ is the element of channel matrix $\mathbf{H}$. In addition, $x_{k \to m}^v(\tau)$ and $v_{k \to m}^v(\tau)$ denote the mean and variance passing from the $k$th variable node to $m$th sum node respectively, $x_{m \to k}^s(\tau)$ and $v_{m \to k}^s(\tau)$ denote the mean and variance passed from $m$th sum node to $k$th variable node respectively. The initial value $\mathbf{v}^v(0)$ equals to $+\boldsymbol{\infty}$ and $\mathbf{x}^v(0)$ equals to $\mathbf{0}$, where $\mathbf{v}^v(\tau)$ and $\mathbf{x}^v(\tau)$ are vectors containing $v_{k\to m}^v(\tau)$ and $x_{k\to m}^v(\tau)$, respectively.

\subsection{Message Update at Variable Nodes of GMPID}
The message update at the variable nodes is denoted by\vspace{-0.3cm}
\begin{equation}\label{var_mes}
\!\!\!\!\!\left\{\!\!\!\! {\begin{array}{*{20}{l}}
{v_{k \to m}^v(\tau) = {{( {\sum\limits_{i\ne m} {h_{ik}^2v_{i \to k}^{s{\;^{ - 1}}}(\tau) + \bar{v}_{{k}}^{ l^{-1}}\;} } )}^{ - 1}},}\\
{x_{k \to m}^v(\tau) \!=\! v_{k \to m}^v(\tau)(\!\!\sum\limits_{i \ne m}\!\! {{h_{ik}}v_{i \to k}^{s{\;^{ - 1}}}(\tau)x_{i \to k}^s(\!\tau\!)}\! +\! \bar{v}_{{k}}^{ l^{-1}}\!\bar{x}_k^{l}) }.
\end{array}} \right.\vspace{-0.2cm}
\end{equation}
where $k \in \mathcal{N}_u, i,m \in \mathcal{N}_r$ and $\bar{v}_{{k}}^{ l}$ and $\bar{x}_k^{l}$ denote the estimated variance and mean of $x_k$ from the decoder $k$. Unlike the GMPID in \cite{Lei2015}, here the messages $\bar{v}_i^l$ and $\bar{x}_i^l$, $i\in \mathcal{N}_u$ from the user decoders are considered in the variable update.

\subsection{Decision and Extrinsic Output of GMPID}
The decision is made based on all the messages coming from all the sum nodes. When the MSE of the GMPID meet the requirement or the number of iteration reaches the limit, we output \vspace{-0.2cm}
\begin{equation}\label{np3}
\left\{ \begin{array}{l}
\hat{v}_{k} = {(\, {\sum\limits_{m } {h_{mk}^2v_{m \to k}^{s{\;^{ - 1}}}(\tau)\,} }+\bar{v}_{{k}}^{ l^{-1}} )^{ - 1}},\\
\hat{x}_{k} = \hat{v}_{k}(\sum\limits_{m } {h_{mk}v_{m \to k}^{s{\;^{ - 1}}}(\tau)x_{m \to k}^s(\tau) + \bar{v}_{{k}}^{ l^{-1}}\bar{x}_k^{l}})\;,
\end{array} \right.\vspace{-0.3cm}
\end{equation}
for $k \in \mathcal{N}_u, m \in \mathcal{N}_r$. However, during the iteration, the GMPID output the extrinsic messages by \vspace{-0.2cm}
\begin{equation}\label{np2}
\left\{ \begin{array}{l}
\bar{v}^e_{k} = {(\, {\sum\limits_{m } {h_{mk}^2v_{m \to k}^{s{\;^{ - 1}}}(\tau)\,} } )^{ - 1}},\\
\bar{x}_{k}^e = \bar{v}^e_{k}\sum\limits_{m } {h_{mk}v_{m \to k}^{s{\;^{ - 1}}}(\tau)x_{m \to k}^s(\tau)\;\;},
\end{array} \right.\vspace{-0.2cm}
\end{equation}
where $k \in \mathcal{N}_u, m \in \mathcal{N}_r$.

\subsection{Complexity of GMPID}
As the variance calculations are independent of the received signals $\textbf{\emph{y}}$ and the means, it can be pre-computed before the iteration. In each iteration, it needs about $4N_uN_r$ multiplications. Therefore, the complexity is as low as $\mathcal{O}(N_uN_rN_{ite}^{ese}N_{ite}^{out})$, where $N_{ite}^{ese}$ is the number of iterations at ESE.

\subsection{Variance Convergence of GMPID}
The following gives the variance convergence of GMPID.

\textbf{\textit{Theorem 1:}} \emph{In the symmetric large-scale MIMO-NOMA systems, i.e., $\bar{v}_i^l=\bar{v}^l$ for $\forall i\in \mathcal{N}_u$, $\beta= N_u/N_r$ is fixed and $N_u\to \infty$, the variances of GMPID converge to that of the LMMSE detection.}

\begin{IEEEproof}
From (\ref{np1}) and (\ref{var_mes}), we have
\vspace{-0.2cm}\begin{equation}\label{p1}
\!\!\!\!\!\!\!{v_{k \to m}^v(\tau) \!=\!\! {{\Big( \!{\sum\limits_{i\ne m} {h_{ik}^2{{( {\sum\limits_{j } {h_{ij}^2v_{j \to i}^v(\tau-1) \!+\! \sigma _n^2\;} } )}^{ - 1}} \!\!\!\!+ \bar{v}_k^{l^{-1}}\;} } \!\!\!\Big)}^{\!\!\! - 1}}}\!\!\!\!\!.\vspace{-0.2cm}
\end{equation}
As the initial value $\mathbf{v}^v(0)$ is equal to $+\boldsymbol{\infty}$, it is easy to see that $\mathbf{v}^v(\tau) > 0$ for any $\tau>0$ during the iteration. So $\mathbf{v}^v(\tau)$ has a lower bound $\mathbf{0}$. From (\ref{p1}), we can see that $\mathbf{v}^v(\tau)$ is a monotonically non-increasing function with respect to $\mathbf{v}^v(\tau-1)$. Moreover, we can get $\mathbf{v}^v(1)<\mathbf{v}^v(0)=+\boldsymbol{\infty}$ for the first iteration. The inequations correspond to the component-wise inequality. Therefore, it can be shown that $\mathbf{v}^v(\tau)\leq \mathbf{v}^v(\tau-1)$ with $\mathbf{v}^v(1) \leq \mathbf{v}^v(0)$ from the monotonicity of the iteration function. This means that $\{\mathbf{v}^v(\tau)\}$ is a monotonic decreasing sequence but is lower bounded. Thus, sequence $\{\mathbf{v}^v(\tau)\}$ converges to a certain value, i.e., $\mathop {\lim }\limits_{\tau \to  \infty } {\mathbf{v}^v}(\tau)=\mathbf{v}^*$.

With the symmetry ($\bar{v}_i^l=\bar{v}^l$ for $\forall i\in \mathcal{N}_u$) of all the elements of $\mathbf{v}^*$, we can get $v_{k\to m}^*=\hat{v}, i\in \mathcal{N}_u\,$, $m\in \mathcal{N}_r$. Thus, from  (\ref{p1}), the convergence point $\hat{v}$ can be solved by\vspace{-0.15cm}
\begin{equation}\label{p2}
{\hat{v} = {{\Big( {\sum\limits_{i\ne m } {h_{ik}^2{{(\hat{v} {\sum\limits_{j } {h_{ij}^2 + \sigma_n^2} } )}^{ - 1}} + \bar{v}^{l^{-1}}} } \Big)}^{ - 1}}}.\vspace{-0.15cm}
\end{equation}
It can be rewritten as
\begin{equation}\label{p3}
\bar{v}^{l^{-1}} \!\!\sum\limits_{j } \!{h_{ij}^2} {{\hat v }^2} + (\sigma _n^2\bar{v}^{l^{-1}} + \sum\limits_{i\ne m} {h_{ik}^2\; - \sum\limits_{j} {h_{ij}^2\;} } ){{\hat v }}\! -\! \sigma _n^2 = 0.\vspace{-0.15cm}
\end{equation}
When $N_r$ is large, taking an expectation on (\ref{p3}) with respect to the channel parameters $h_{ik}^2$ and $h_{ij}^2$, we get\vspace{-0.15cm}
\begin{equation}\label{p4}
N_u\bar{v}^{l^{-1}}{{\hat v }^2} + (\sigma _n^2\bar{v}^{l^{-1}} + N_r - N_u ){{\hat v }} - \sigma _n^2 = 0.\vspace{-0.15cm}
\end{equation}
Then $\hat{v}$ is the positive solution of (\ref{p4}), i.e.,
\begin{equation}\label{p42}
\!\!\!\!\!\!\hat{v} =\!\! \frac{{\sqrt {{{\!(\sigma _n^2\bar{v}^{l^{-1}}\!\! \!+ \!N_r\!\! -\!\! N_u)}^2}\!\!\! +\! 4N_u \bar{v}^{l^{-1}}\!\!\sigma _n^2} \! - \!(\sigma _n^2\bar{v}^{l^{-1}} \!\!\!\!+ \!N_r\!\! - \!\!N_u)}}{{2N_u \bar{v}^{l^{-1}}}}.
\end{equation}
It is easy to verify that $\hat{v}=\hat{v}_{mmes}$ (refer to \emph{Proposition 1}).
\end{IEEEproof}

Similarly, sequence $\{\mathbf{v}^s(\tau)\}$ converges to a certain value, i.e., $v^s_{m\to k} \to v^s$, $k \in \mathcal{N}_u$ and $m \in \mathcal{N}_r$. From (\ref{np1}), we can get \vspace{-0.2cm}
\begin{equation}\label{p8}
v^s \approx N_u{\hat v} + \sigma _n^2, \; and \; \gamma  ={{{\hat v}} \mathord{\left/
 {\vphantom {{{\hat v}} {{v^s}}}} \right.
 \kern-\nulldelimiterspace} {{v^s}}} = \frac{1}{{{{N_u}} + {{\sigma _n^2} \mathord{\left/
 {\vphantom {{\sigma _n^2} {{{\hat v }}}}} \right.
 \kern-\nulldelimiterspace} {{{\hat v }}}}}} .
\end{equation}

\subsection{Mean Convergence of GMPID}
Unlike the variances, the means do not always converge. Firstly, we give the classical iterative algorithms \cite{Axelsson1994} by\vspace{-0.15cm}
\begin{equation}\label{a1}
{\textbf{\emph{x}}}(t) = \mathbf{B}{\textbf{\emph{x}}}(t - 1) + {\textbf{\emph{c}}},\vspace{-0.15cm}
\end{equation}
where $\mathbf{B}$ is a given matrix, and ${\textbf{\emph{c}}}$ is a given vector. The \emph{Jacobi detector} and \emph{Richardson algorithm} are special cases of it.

\textbf{\textit{Proposition 2 }}\cite{Axelsson1994}: \emph{Assuming that the matrix $\mathbf{I}-\mathbf{B}$ is invertible, the iteration (\ref{a1}) converges to the exact solution ${\textbf{{x}}}^*=(\mathbf{I}-\mathbf{B})^{-1}\mathbf{c}$ for any initial guess ${\textbf{{x}}}\left(0\right)$ if $\mathbf{I}-\mathbf{B}$ is strictly (or irreducibly) diagonally dominant or $\rho \left( \mathbf{B }\right) < 1$, where $\rho(\mathbf{B})$ is the spectral radius of $\mathbf{B}$.}

The mean convergence of GMPID are given as following.

\textbf{\textit{Theorem 2:}} \emph{When $\beta= N_u/N_r$ is fixed and $N_u\to \infty$, the GMPID converges to\vspace{-0.15cm}
\begin{equation}\label{f_gm}
\hat{\mathbf{x}}=\left( \theta \mathbf{H}^T\mathbf{H}+ I_{N_u} \right)^{ - 1}\left( \theta \mathbf{H}^T\textbf{\emph{y}}+ \alpha\bar{\mathbf{x}}^l\right),\vspace{-0.15cm}
\end{equation}
where $\theta=\hat v/ \sigma^2_n$ and $\alpha=\hat{v}/\bar{v}^l$, if any of the following conditions holds.}

\emph{1. The matrix $\mathbf{I}_{N_r} + \gamma\left({\mathbf{H}}\mathbf{H}^T-\mathbf{D}_{{\mathbf{H}}\mathbf{H}^T}\right)$ is strictly or irreducibly diagonally dominant,}

\emph{2. $\rho \left( {\gamma ({\mathbf{H}}\mathbf{H}^T - {\mathbf{D}_{{\mathbf{H}}\mathbf{H}^T}})} \right) < 1$, where  $\gamma=\hat{v}/v^s$}.

\begin{IEEEproof}
From (\ref{np1}) and (\ref{var_mes}), we have
\vspace{-0.15cm}\begin{equation*}\label{p9}
\!\!x_{m\to k}^s(\tau+1)\!=\!y_m\!-\!\!\sum\limits_{i}\!{h_{mi}v_{i\to m}^v\!(\!\tau\!)}\!\Big( \!\!\sum\limits_{j\ne m}\!\!h_{ji}v_{j\to i}^{s^{-1}}(\!\tau\!)x_{j\to i}^s(\!\tau\!) + \bar{v}_i^{l^{-1}}\!\!\!\bar{x}_i^l\!\Big)\vspace{-0.25cm}
\end{equation*}
From the variance convergence analysis, the $v^s_{m\to k}(\tau)$ and $v^v_{k\to m}(\tau) $ converge to $v^s$ and $\hat v$. Therefore,\vspace{-0.15cm}
\begin{equation}\label{p10}
x_{m\to k}^s(\tau)=y_m-\sum\limits_{i}{h_{mi}}( \gamma\sum\limits_{j\ne m}h_{ji} x_{j\to i}^s(\tau-1) + \alpha\bar{x}_i^l),\vspace{-0.25cm}
\end{equation}
where $\alpha=\hat v/\bar{v}^l$. Then, we get $x_{k \to m}^v(\tau)=x^v_{k}(\tau)$. Thus, \vspace{-0.1cm}
\begin{equation}\label{p11}
\mathbf{x}^s(\tau) = \textbf{\emph{y}} -  \gamma (\mathbf{H}\mathbf{H}^T-\mathbf{D}_{\mathbf{H}{\mathbf{H}^T}}) \mathbf{x}^s(\tau-1)-\alpha \mathbf{H}\bar{\mathbf{x}}^l,\vspace{-0.1cm}
\end{equation}
where  $\mathbf{x}^v(\tau)={\left[ {{{x}_1^v}(\tau)\;{{x}_2^v}(\tau)\; \cdots \;{{x}_{N_u}^v}(\tau)} \right]^T}$, $\mathbf{D}_{\mathbf{H}\mathbf{H}^T}=diag\{d_{11},d_{22}, \cdots ,d_{N_uN_u}\}$ is a diagonal matrix and $d_{kk}=\mathbf{h}_k^T\mathbf{h}_k$, {\small$k\in \{1,2,\cdots,N_u\}$} are the diagonal elements of the matrix $\mathbf{H}\mathbf{H}^T$. When $N_u$ is large, from the law of large numbers, the matrix $\mathbf{D}_{\mathbf{H}\mathbf{H}^T}$ can be approximated by $K\mathbf{I}_{N_r}$. Assuming that $\{\mathbf{x}^s(\tau)\}$ converges to $\mathbf{x}^*$, then we have\vspace{-0.1cm}
\begin{equation}\label{p12}
{\mathbf{x}^*} = {\left( {(1-\gamma N_u){\mathbf{I}_{N_r}} +\gamma\mathbf{H}{\mathbf{H}^T}} \right)^{ - 1}}(\textbf{\emph{y}}-\alpha \mathbf{H} \bar{\mathbf{x}}^l).\vspace{-0.1cm}
\end{equation}
From (\ref{np3}), we the full estimation as\vspace{-0.1cm}
\begin{equation}\label{g_fm1}
{\hat{\mathbf{x}}} = \gamma \mathbf{H}^T\mathbf{x}^*+\alpha \bar{\mathbf{x}}^l.\vspace{-0.1cm}
\end{equation}
That is the GMPID converges to $\hat{\mathbf{x}}$ if it converges. In addition,\vspace{-0.1cm}
 \begin{eqnarray}\label{g_fm2}
{\hat{\mathbf{x}}}& = &\!\!\!\!\!\left( \theta \mathbf{H}^T\mathbf{H}+ I_{N_u} \right)^{ - 1}\theta \mathbf{H}^T\textbf{\emph{y}}+\alpha \left( \theta \mathbf{H}^T\mathbf{H}+ I_{N_u} \right)^{ - 1}\bar{\mathbf{x}}^l\nonumber\\
&=& \!\!\!\!\!\left( \theta \mathbf{H}^T\mathbf{H}+ I_{N_u} \right)^{ - 1}\left( \theta \mathbf{H}^T\textbf{\emph{y}}+ \alpha \bar{\mathbf{x}}^l\right),\vspace{-0.25cm}
\end{eqnarray}
where $\theta=\hat v/ \sigma^2_n=(\gamma^{-1}-N_u)^{-1}$, the first equation is based on the \emph{matrix inverse lemma}. Let $ \textbf{\emph{c}}= \textbf{\emph{y}}-\alpha \mathbf{H}\bar{\mathbf{x}}^l$ and $\mathbf{B}= -  \gamma (\mathbf{H}\mathbf{H}^T-\mathbf{D}_{\mathbf{H}{\mathbf{H}^T}})$, then (\ref{p11}) is a classical iterative algorithm \cite{Axelsson1994}. Thus, we can get Theorem 2 with Proposition 2.
\end{IEEEproof}

When $N_u\rightarrow \infty$, from random matrix theory \cite{verdu1998}, we have \vspace{-0.15cm}
\begin{equation}\label{radius1}
\rho_{_{GMPID}} \!\!=\!\! \rho(\gamma(\mathbf{H}\mathbf{H}^T\!\!-\mathbf{D}_{\mathbf{H}\mathbf{H}^T}))\to \gamma N_u(\beta^{-1}\!+2\sqrt{\beta^{-1}}).
\end{equation}
Then, from the Theorem 2, we have the following corollary.

\textbf{\emph{Corollary 1:}} \emph{When $\beta= N_u/N_r$ is fixed and $N_u\to \infty$, the GMPID converges to \vspace{-0.1cm}
\begin{equation}
\hat{\mathbf{x}}=\left( \theta \mathbf{H}^T\mathbf{H}+ I_{N_u} \right)^{ - 1}\left( \theta \mathbf{H}^T\textbf{\emph{y}}+ \alpha\bar{\mathbf{x}}^l\right),\vspace{-0.1cm}
\end{equation}
if $\gamma(N_r+2\sqrt{N_rN_u})<1$,
where $\gamma=(N_u+\sigma_n^2/\hat{v})^{-1}$, and $\hat{v}$ is given in (\ref{p42}). This sufficient condition can be simply approximated to\vspace{-0.25cm}
\begin{equation}
\beta >(\sqrt{2}-1)^{-2}.\vspace{-0.5cm}
\end{equation}}

Different from the case $N_u<N_r$, the GMPID does not converge to the LMMSE detection even if the GMPID is convergent when $N_u>N_r$.

\textbf{\emph{Corollary 2:}} \emph{Even if the GMPID is convergent, it has a worse MSE performance than that of the MMSE detection.}

\begin{IEEEproof}
The proof is omitted due to the page limit.
\end{IEEEproof}

\section{A New fast-convergence Detector SA-GMPID}
As shown in the convergence analysis in Section IV, the GMPID does not converge to the optimal LMMSE detection. The main reason is that the spectral radius of GMPID does not achieve the minimum value. Therefore, we propose a new \emph{scaled-and-added} GMPID (SA-GMPID) by modifying the GMPID with linear operators, that is \emph{i)} keeping the variance update unchanged as it converges to the LMMSE detection, \emph{ii)} scaling the received $\mathbf{y}$ and the channel matrix $\mathbf{H}$, i.e., ${\mathbf{H}'}=\sqrt{w}\mathbf{H}$ and $\textbf{\emph{y}}'=\sqrt{w}\textbf{\emph{y}}$, where $h'_{mk}=\sqrt{w}h_{mk}$ is an element of matrix $\mathbf{H}'$ and $w$ is a relaxation parameter, and \emph{iii)} adding a new term $\!(\!w\!-\!1\!)x_{m \to k}^s\!(\!\tau\!-\!1\!)$ for the mean message update at the sum node. Then, we can optimize the relaxation parameter $w$ to minimize the spectral radius of SA-GMPID.

\subsection{Message Update at Sum Nodes of SA-GMPID}
The message update at the sum nodes (\ref{np1}) is modified as
\begin{equation}\label{FC2}
\!\!\!\!\!\!\!\left\{ \!\!\!\!\begin{array}{l}
x_{m \to k}^s(\!\tau\!) \!=\! \!{y'_m} \!\!\!-\!\!\! \sum\limits_{i}\! {h'_{mi}x_{i \to m}^v(\!\tau\!-\!1\!)}\!\!-\!\!(w\!-\!1)x_{m \to k}^s(\tau\!\!-\!\!1), \\
v_{m \to k}^s(\tau)= \sum\limits_{i} {h_{mi}^2v_{i \to m}^v(\tau-1)\! +\! \sigma _n^2\;},
\end{array} \right.\vspace{-0.15cm}
\end{equation}
for $i, k \in \mathcal{N}_u,m \in \mathcal{N}_r$.\vspace{-0.15cm}
\subsection{Message Update at Variable Nodes of SA-GMPID}
The message update of the variable nodes (\ref{var_mes}) is changed to
\begin{equation}\label{FC1}
\!\!\!\left\{ \!\!\!{\begin{array}{*{20}{l}}
{v_{k \to m}^v(\tau) = {{( {\sum\limits_{i\ne m} {h_{ik}^2v_{i \to k}^{s{\;^{ - 1}}}(\tau) + \bar{v}_{{k}}^{ l^{-1}}\;} } )}^{ - 1}},}\\
{x_{k \to m}^v(\tau) = \bar{v}_{k}^l(\sum\limits_{i \ne m} {{h'_{ik}}\bar{v}_{i }^{s{\;^{ - 1}}}x_{i \to k}^s(\tau)} + \bar{v}_{{k}}^{ l^{-1}}\bar{x}_k^{l}) },
\end{array}} \right.\quad\vspace{-0.2cm}
\end{equation}
where $\bar{v}_m^s= \sum\limits_{k} {h_{mk}^2\bar{v}_{k}^l + \sigma _n^2}$, and $k \in \mathcal{N}_u, i,m \in \mathcal{N}_r$.\vspace{-0.2cm}

\subsection{Decision and Extrinsic Output of SA-GMPID}
The decision of the SA-GMPID is given by
\begin{equation}\label{FC4}
\left\{ \begin{array}{l}
{ \hat{v}_{k} = {(\, {\sum\limits_{m } {h_{mk}^2v_{m \to k}^{s{\;^{ - 1}}}(\tau)\,} }+\bar{v}_{{k}}^{ l^{-1}} )^{ - 1}},}\\
\hat{x}_{k} = \bar{v}^l_{k}(\sum\limits_{m } {h'_{mk}\bar{v}_{m}^{s{\;^{ - 1}}}x_{m \to k}^s(\tau) + \bar{v}_{{k}}^{ l^{-1}}\bar{x}_k^{l}})\;,
\end{array} \right.\quad\vspace{-0.2cm}
\end{equation}
for $k \in \mathcal{N}_u, m \!\in\! \mathcal{N}_r$. The extrinsic messages of SA-GMPID is \vspace{-0.15cm}
\begin{equation}\label{FC3}
\left\{ \begin{array}{l}
{\bar{v}^e_{k} = {(\, {\sum\limits_{m } {h_{mk}^2v_{m\to k}^{s{\;^{ - 1}}}(\tau)\,} } )^{ - 1}},}\\
\bar{x}_{k}^e = {(\bar{v}_k^l+\bar{v}^e_{k})}\sum\limits_{m } {h'_{mk}v_{m }^{s{\;^{ - 1}}}x_{m \to k}^s(\tau)}+\bar{x}_k^l,
\end{array} \right.\vspace{-0.2cm}
\end{equation}
where $\bar x_k^e = \bar v_k^e(\hat v_k^{ - 1}{{\hat x}_k} - \bar v_k^{{l^{ - 1}}}\bar x_k^l)$, and $k \in \mathcal{N}_u, m \in \mathcal{N}_r$.

\textbf{\emph{Remark 1}}: From Theorem 2, we can see that the GMPID converges to the LMMSE if it is convergent with $\hat{v}=\bar{v}^l$. Therefore, we let $v_{k\to m}^v(\tau)=\bar{v}_k^l$ and $v_{m\to k}^s(\tau)=\bar{v}_m^s$ in the mean message update (\ref{FC1}), to assure that the SA-GMPID can converge to the LMMSE detection.

As the variance expressions are the same as the GMPID, the variances of SA-GMPID converge to the same results as the GMPID. Hence, we focus on the means of the SA-GMPID.

\subsection{Mean Convergence of SA-GMPID}
Similarly, we get the matrix form of SA-GMPID
\begin{equation}\label{FC8}
\mathbf{x}^s(\!\tau\!) \!=\! \textbf{\emph{y}}' \!\!-\!  \!\left[\tilde{\gamma} (\mathbf{H}'\mathbf{H}'^T \!\!\!-\!\mathbf{D}_{\mathbf{H}'{\mathbf{H}'^T}}) \!-\!(w\!-\!1)\mathbf{I}_{N_r}\!\right] \!\mathbf{x}^s(\tau\!-\!1\!)\!-\! \mathbf{H}'\bar{\mathbf{x}}^l .\vspace{-0.2cm}
\end{equation}
Then, we can have the following theorem.

\textbf{\emph{Theorem 3:}}  \emph{When $\beta= N_u/N_r$ is fixed and $N_u\to \infty$, the SA-GMPID converges to the LMMSE estimation if the relaxation parameter $w$ satisfies $0<w<2/\lambda_{max}^\mathbf{A}$, where $\lambda_{max}^\mathbf{A}$ is the largest eigenvalue of matrix $\mathbf{A}=\tilde{\gamma}\left(\mathbf{H}\mathbf{H}^T-\mathbf{D}_{\mathbf{H}\mathbf{H}^T}\right) +\mathbf{I}_{N_r}$, $\tilde{\gamma}  = {({N_u} + \sigma_n^2/\bar{v}^l)^{ - 1}}$.}

\begin{IEEEproof}
The proof is omitted due to the page limit.
\end{IEEEproof}

The relaxation parameter $w$ can be optimized by \cite{Gao2014} \vspace{-0.15cm}
\begin{equation}\label{FC10}
w^* =2/({\lambda}_{min}^\mathbf{A}+{{\lambda}_{max}^\mathbf{A}}).\vspace{-0.15cm}
\end{equation}
It minimizes the spectral radius of $\mathbf{I}_{N_r}-w\mathbf{A}$ to $\rho_{min}(\mathbf{I}_{N_r}-w\mathbf{A})= \frac{{\lambda}_{max}^\mathbf{A}-{\lambda}_{min}^\mathbf{A}}{{\lambda}_{max}^\mathbf{A}+{\lambda}_{min}^\mathbf{A}}<1$ . Furthermore, we have\vspace{-0.15cm}
  \begin{equation}
 w^* = {1 \mathord{\left/
 {\vphantom {1 {\left( {1 + \tilde{\gamma} M \beta} \right)}}} \right.
 \kern-\nulldelimiterspace} {\left( {1 + \tilde{\gamma} N_u \beta^{-1}} \right)}}=1/(1+\tilde{\gamma} N_r),\vspace{-0.15cm}
 \end{equation}
 and the minimal spectral radius of SA-GMPID
 \begin{equation}\label{FC12}
 \rho_{_{SA-GMPID}} = \rho_{min}(\mathbf{I}_{K}-w\mathbf{A})=\frac{2\tilde{\gamma}\sqrt{N_uN_r}}{1+\tilde{\gamma} N_r}<1. \vspace{-0.2cm}
 \end{equation}
Comparing (\ref{FC12}) with (\ref{radius1}), we have ${\rho_{_{GMPID}}}/{\rho_{_{SA-GMPID}}}>1$. Therefore, we have the following corollary.

\textbf{\emph{Corollary 3:}} \emph{The proposed SA-GMPID converges faster than the GMPID when $\beta= N_u/N_r$ is fixed and $N_u\to \infty$.}

It should be noted that the complexity of SA-GMPID algorithms is almost the same as that of the GMPID. \vspace{-0.2cm}

\section{Simulation Results}
In this section, we present the numerical simulation results for the MIMO-NOMA systems. Assume that the sources are i.i.d. with $x_k\sim\mathcal{N}(0,1)$, $SNR=\frac{1}{\sigma^2_n}$.
\begin{figure}[t]\vspace{-0.4cm}
  \centering
  \includegraphics[width=9cm]{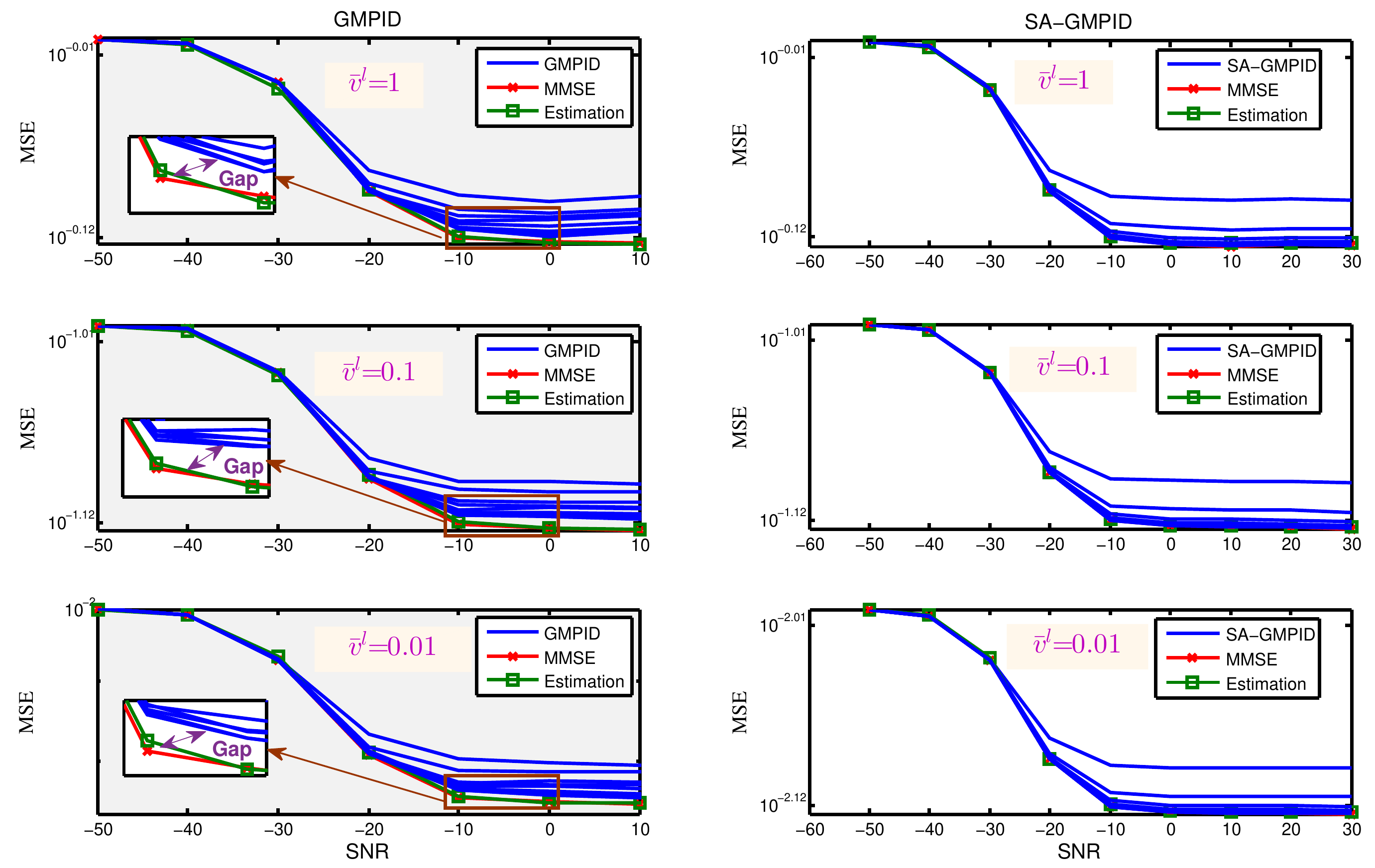}\\\vspace{-0.3cm}
  \caption{MSE performance comparison between the MMSE detection, GMPID and SA-GMPID with $1\sim 10$ iterations; the performance estimate of the MMSE detection and the proposed GMPID. $N_u=400$, $N_r=100$, $\beta=4$ and $\bar{v}^l=[1, \; 0.1,\; 10^{-2}]$. }\label{f5}\vspace{-0.2cm}
\end{figure}

\begin{figure}[t]\vspace{-0.0cm}
  \centering
  \includegraphics[width=8.0cm]{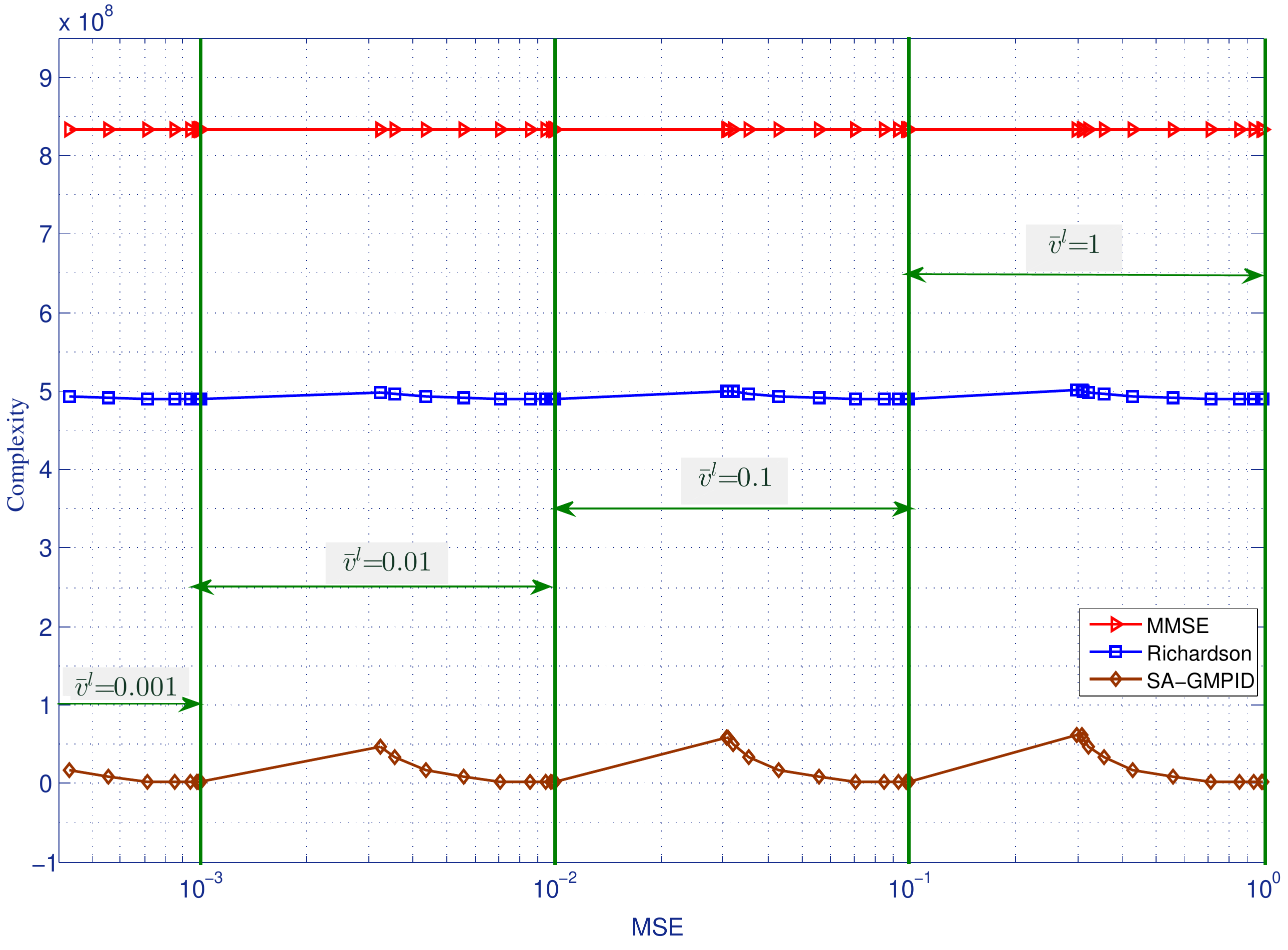}\\\vspace{-0.4cm}
  \caption{Complexity comparison between the MMSE detection, Richardson algorithm and SA-GMPID for $N_u=1000$, $N_r=700$, $\beta=10/7$, and $\bar{v}^l=[1, \; 0.1,\; 10^{-2},\; 10^{-3}]$. }\label{f7}\vspace{-0.45cm}
\end{figure}
Fig. \ref{f5} presents average MSE performance comparison between the MMSE detection, GMPID and SA-GMPID with $1\sim 10$ iterations, where $N_u=400$, $N_r=100$, $\beta=4$ and $\bar{v}^l=[1, \; 0.1,\; 10^{-1},\; 10^{-2},\; 10^{-3}]$. The performance estimate of the MMSE detection (\ref{PA6}) is also shown in Fig. \ref{f5}. It can be seen that for the different input $\bar{v}^l$, the performance estimate of the MMSE detection (\ref{PA6}) is accurate. From the left three subfigures in Fig. \ref{f5}, we can see that the GMPID converges to a higher MSE than that of the MMSE detection for any input $\bar{v}^l$, which coincides with the conclusions in Theorem 2. However, from the right three subfigures in Fig. \ref{f5}, it can be seen that the proposed SA-GMPID always converges to the MMSE detection and has a faster convergence speed than the GMPID, which verifies the conclusions in Theorem 3.

Fig. \ref{f7} illustrates the complexity comparison between the different detection algorithms with different MSE for $1000\times700$ MIMO-NOMA system with $\beta=10/7$ for the different input $\bar{v}^l$. We can see that the complexity of the SA-GMPID detector increases with decreasing the MSE. It also shows that  the proposed SA-GMPID has a good convergence performance and much lower computational complexity. The MSE performance of GMPID, Jacobi algorithm and GaBP algorithm are divergent in this case as the $\beta=10/7$ is very close to 1.

\begin{figure}[t]\vspace{-0.4cm}
  \centering
  \includegraphics[width=8.5cm]{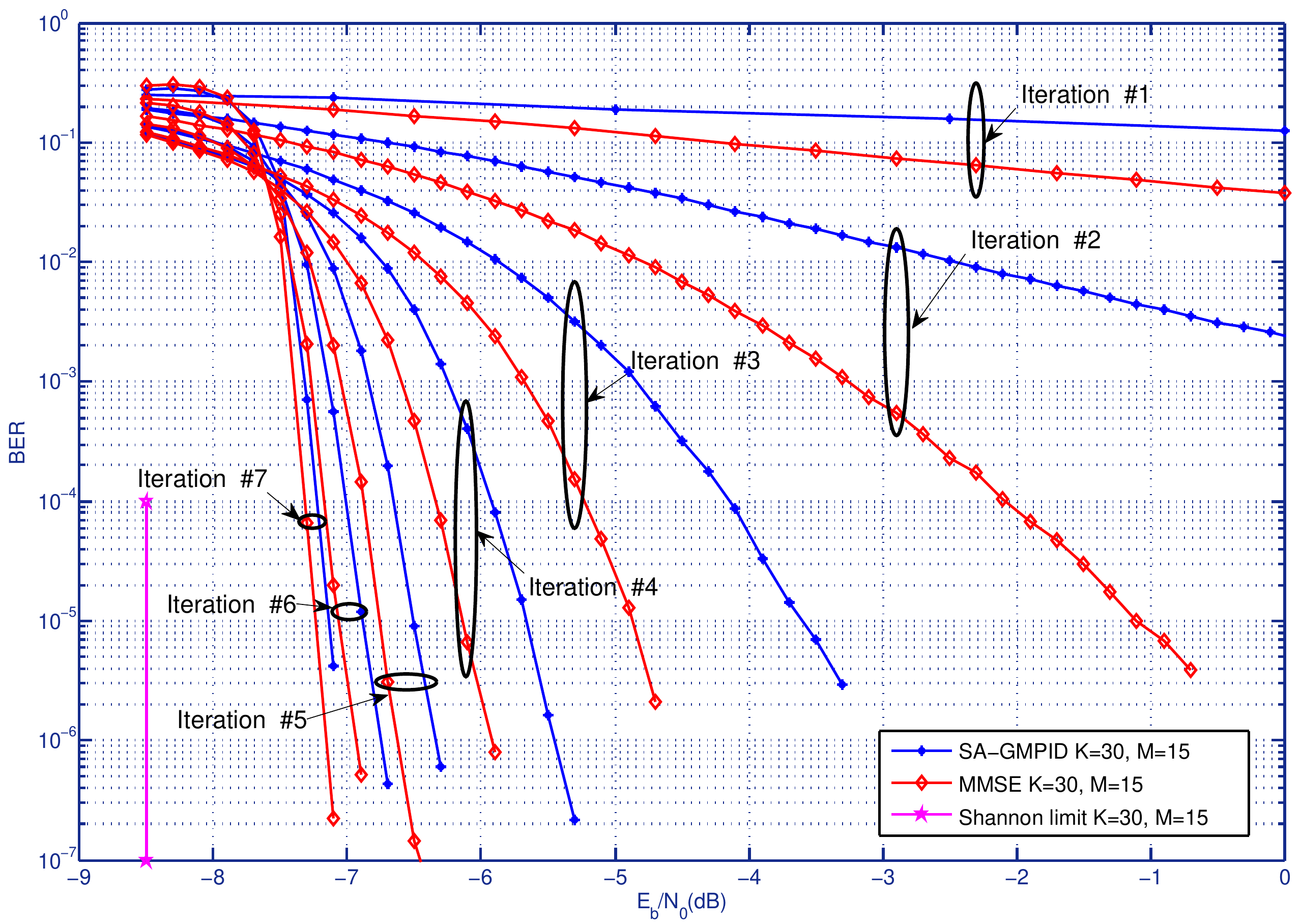}\\\vspace{-0.3cm}
  \caption{BER performances of SA-GMPID and MMSE detection with Gaussian approximation for the discrete MIMO-NOMA systems, where $K=30$, $M=15$, $\beta=2$ and $N_{ite}^{ese}=1$, $N_{ite}^{out}=1\sim7$. }\label{f8}\vspace{-0.4cm}
\end{figure}

Fig. \ref{f8} presents the BER of the SA-GMPID and MMSE detection for a practical $30\times15$ MIMO-NOMA systems transmitting digital modulation waveforms with $\beta=2$. The original GMPID is divergent in this system for the small value of $\beta$. In this system, each user is encoded with a Turbo Hadamard channel code \cite{Ping2003_2}, where the code length is $2.82\times 10^5$. A 10-bit superposition coded modulation \cite{Yuan2014} is employed for each user to produce Gaussian like transmitting signals. Hence, the transmitting length of each user is $2.82\times 10^4$, the rate of each user is $R_u=0.1452$ bits/symbol, and the system sum rate is 4.356 bits per channel use. $E_b/N_0$ is calculated by $E_b/N_0=\frac{1}{2R_u\sigma^2_n}$. The Shannon limit of $30\times15$ MIMO-NOMA system is $E_b/N_0=-8.523$dB. In this simulation, we let $N_{ite}^{ese}=1$, which decreases the complexity of SA-GMPID to $\mathcal{O}(N_uN_rN_{ite}^{out})$. Fig. \ref{f8} shows that for the $30\times10$ MIMO-NOMA system, it only needs about 7 iterations for the proposed SA-GMIPD converging to the MMSE detection, and the performance is only 1.5dB away from the Shannon limit. It consists with our theoretic analysis that the proposed SA-GMIPD can always converge to the MMSE detection quickly, even when $\beta=2<(\sqrt{2}-1)^{-2}$.\vspace{-0.15cm}

\section{Conclusion}
We consider the GMPID for the MIMO-NOMA with massive access that the number of users $N_u$ is larger than the number of antennas $N_r$.The convergence of GMPID is analysed in this paper. It is proved that the variances of GMPID converge to the MSE of MMSE detection. Two sufficient conditions that the GMPID is convergent with a higher MSE than that of the MMSE detection are presented. A new SA-GMPID algorithm is proposed, which is proved to converge to the MMSE detection in mean and variance for any $N_u>N_r$. Numerical results are provided to verify the proposed theoretical results.

%\newpage

\end{document}